\documentclass[aps,prb,twocolumn,groupedaddress,amsfonts,showpacs]{revtex4}
\usepackage{graphicx,color}
\usepackage{bm}

\begin{document}
\title{Angular dependence of the tunneling anisotropic magnetoresistance}
\author{A. Matos-Abiague, M. Gmitra, and J. Fabian}
\affiliation{Institute for Theoretical Physics, University of
Regensburg, 93040 Regensburg, Germany}
\date{\today}

\begin{abstract}
Based on general symmetry considerations we investigate how the
dependence of the tunneling anisotropic magnetoresistance (TAMR)
on the magnetization direction is determined by the specific form
of the spin-orbit coupling field. By extending a phenomenological
model, previously proposed for explaining the main trends of the
TAMR in (001) ferromagnet/semiconductor/normal-metal magnetic
tunnel junctions (MTJs) [J. Moser {\it et al.}, Phys. Rev. Lett.
99, 056601 (2007)], we provide a unified qualitative description
of the TAMR in MTJs with different growth directions. In
particular, we predict the forms of the angular dependence of the
TAMR in (001),(110), and (111) MTJs with structure inversion
asymmetry and/or bulk inversion asymmetry. The effects of in-plane
uniaxial strain on the TAMR are also investigated.
\end{abstract}


\maketitle

\section{Introduction}

The tunneling anisotropic magnetoresistance (TAMR) effect refers
to the dependence of the magnetoresistance of magnetic tunnel
junctions (MTJs) on the absolute orientation(s) of the
magnetization(s) in the ferromagnetic lead(s) with respect to the
crystallographic
axes.\cite{Gould2004:PRL,Brey2004:APL,Ruster2005:PRL,Saito2005:PRL}
Unlike the conventional tunneling magnetoresistance (TMR) effect,
the TAMR is not only present in MTJs in which both electrodes are
ferromagnetic but may also appear in tunneling structures with a
single magnetic electrode.\cite{Gould2004:PRL,Moser2007:PRL}
Because of this remarkable property, if the major challenge of
increasing the size of the effect at room temperature is solved,
the TAMR could be an attractive candidate for applications in the
design of new spin-valve based devices whose components could
operate with a single magnetic lead. In what follows we focus our
discussion on the case of MTJs in which only one of the electrodes
is ferromagnetic.

The TAMR has been experimentally and theoretically investigated in
a variety of systems under different
configurations.\cite{Gould2004:PRL,Brey2004:APL,Ruster2005:PRL,Giddings2005:PRL,
Saito2005:PRL,Shick2006:PRB,Ciorga2007:NJP,Chantis2007:PRL,Gao2007:PRL,Moser2007:PRL,Fabian2007:APS,
Sankowski2007:PRB,Jacob2008:PRB,Khan2008:JPCM,Liu2008:NL,Park2008:PRL,Shick2008:PRB,Matos2009:PRB}
This diversity has made it difficult to build a unified theory of
the TAMR. In fact, although there exists a general consensus in
identifying the spin-orbit coupling (SOC) as the mechanism
responsible for the TAMR, it has been recognized that the way the
SOC influences the TAMR may depend on the considered system and
configuration.

\begin{figure}
\includegraphics[width=6cm]{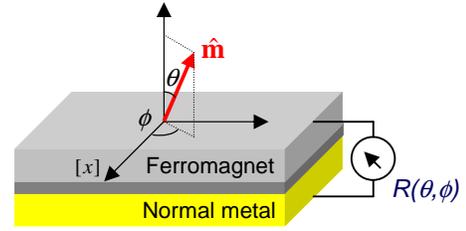}
\caption{Schematics of a MTJ composed of a normal-metal lead
(bottom layer), a semiconductor or insulator barrier (central
layer), and a ferromagnetic electrode (top layer). The vector
$\hat{\mathbf{m}}$ indicates the magnetization orientation, while
$[x]$ denotes a reference crystallographic axis.} \label{syst}
\end{figure}

Two different configurations, the \textit{in-plane} and
\textit{out-of-plane} configurations, have been considered for
investigating the TAMR (for an extensive discussion see
Ref.~\onlinecite{Matos2009:PRB} and references therein). The
in-plane TAMR refers to the changes in the tunneling
magnetoresistance when the magnetization direction, defined with
respect to a fixed reference axis $[x]$, is rotated in the plane
of the ferromagnetic layer. The in-plane TAMR ratio is defined
as\cite{Matos2009:PRB}
\begin{equation}\label{tamr-in-def}
     {\rm TAMR}^{\rm in}_{[x]}(\phi)=
     \frac{R(\theta=90^{\circ},\phi)-R(\theta=90^{\circ},\phi=0)}{R(\theta=90^{\circ},\phi=0)},
\end{equation}
where $R(\theta,\phi)$ denotes the tunneling magnetoresistance for
the magnetization oriented along the direction defined by the unit
vector $\hat{\mathbf{m}}=(\sin\theta \cos\phi,\sin\theta
\sin\phi,\cos\theta)$ (see Fig.~\ref{syst}).

In the out-of-plane configuration, the TAMR measures the changes
in the tunneling magnetoresistance when the magnetization is
rotated within the plane defined by the reference axis $[x]$ and
the direction normal to the ferromagnetic layer. The out-of-plane
TAMR is given by\cite{Matos2009:PRB}
\begin{equation}\label{tamr-out-def}
     {\rm TAMR}^{\rm out}_{[x]}(\theta)=
     \frac{R(\theta,\phi=0)-R(\theta=0,\phi=0)}{R(\theta=0,\phi=0)}.
\end{equation}

An important property of the TAMR is the form of its angular
dependence. It has been experimentally shown that both the
in-plane and out-of-plane TAMR exhibit rather regular, and
relatively simple angular dependence with a well defined symmetry,
in spite of the highly complicated band structure of the
considered
systems.\cite{Saito2005:PRL,Gao2007:PRL,Moser2007:PRL,Park2008:PRL}
This suggests that although the size of the TAMR may depend on the
detailed band structure of the system, its angular dependence is
essentially determined by the symmetry properties of the SOC
field. Here we investigate how the specific form of the TAMR
angular dependence emerges from the properties of the SOC field.

A phenomenological model which incorporate the effects of the
interference of Bychkov-Rashba and Dresselhaus SOCs was recently
developed to explain the in-plane TAMR in (001)
ferromagnet/semiconductor/normal-metal (F/S/NM)
MTJs.\cite{Moser2007:PRL,Fabian2007:APS,Matos2009:PRB} In
particular, it was shown that, in spite of its relative
simplicity, the model was able to reproduce the two-fold symmetric
angular dependence of the in-plane TAMR experimentally observed in
(001) Fe/GaAs/Au MTJs.\cite{Moser2007:PRL,Fabian2007:APS} In such
heterojunctions all the involved materials are cubic in their bulk
form. Therefore, the two-fold anisotropy of the in-plane TAMR must
originate from the interfaces. Here we generalize the model and
provide a unified qualitative description of the angular
dependence of both the in-plane and out-of-plane TAMRs in (001),
(110), and (111) MTJs. We consider systems in which the SOC
originates from structure inversion asymmetry (Bychkov-Rashba-like
SOC) and/or bulk inversion asymmetry (Dresselhaus-like SOC) and
predict new forms of the TAMR angular dependence which could be
tested in future experiments. The effects of uniaxial strain are
also discussed.

\section{Theoretical model}

We consider a MTJ composed of a ferromagnetic electrode and a
normal metal counter-electrode separated by an insulator or a
semiconductor barrier. However, our conclusions are also valid for
the case of MTJs with two ferromagnetic electrodes whose
magnetizations are parallel to each other, since such systems are
qualitatively similar to the case of MTJs with a single
ferromagnetic lead.

The $z$ direction is fixed along the normal to the ferromagnetic
layer (i.e., parallel to the growth direction). The effective
spin-orbit interaction corresponding to the $n$th band can be
written as
\begin{equation}\label{soi}
H_{\rm SO}=\mathbf{w}_{n}(\mathbf{k})\cdot
\mbox{\boldmath$\sigma$},
\end{equation}
where
$\mathbf{w}_{n}(\mathbf{k})=[w_{nx}(\mathbf{k}),w_{ny}(\mathbf{k}),w_{nz}(\mathbf{k})]$
is the effective SOC field associated to the $n$th, $\mathbf{k}$
is the wave vector, and $\mbox{\boldmath$\sigma$}$ is a vector
whose components are the Pauli matrices. Equation (\ref{soi}) is
quite general, since by now we have not considered any specific
form for the SOC field. The detailed form of the SOC field can be
quite complicated as one goes away from the center of the
Brillouin zone and quite different from band to band, as recently
demonstrated by first-principles calculations \cite{Gmitra2009}.

Due to the presence of the spin-orbit interaction, the
transmissivity $T_{n}(\mathbf{k},\hat{\mathbf{m}})$ corresponding
to the $n$th band becomes dependent on the magnetization direction
$\hat{\mathbf{m}}$. Assuming that the strength of the SOC field is
small relative to both the Fermi energy and the exchange
splitting, one can expand the transmissivity in powers of
$\mathbf{w}_{n}(\mathbf{k})$. For a given $n$ and $\mathbf{k}$
there are only two preferential directions in the system, defined
by $\hat{\mathbf{m}}$ and $\mathbf{w}_{n}$. Since the
transmissivity is a scalar function, it can be written, to second
order in the SOC field strength, in the form
\cite{Moser2007:PRL,Fabian2007:APS,Matos2009:PRB}
\begin{eqnarray}\label{tunnel}
    &&T_{n}(\mathbf{k},\hat{\mathbf{m}})\approx
    a_{1n}^{(0)}(\mathbf{k})+a_{1n}^{(1)}(\mathbf{k})[\hat{\mathbf{m}}\cdot\mathbf{w}_{n}(\mathbf{k})]+\nonumber
    \\
    &&
    a_{1n}^{(2)}(\mathbf{k})|\mathbf{w}_{n}(\mathbf{k})|^{2}+a_{2n}^{(2)}(\mathbf{k})
    [\hat{\mathbf{m}}\cdot\mathbf{w}_{n}(\mathbf{k})]^{2},
\end{eqnarray}
which represents the most general expansion (up to second order)
of a scalar function (the transmissivity) in terms of two vectors
($\hat{\mathbf{m}}$ and $\mathbf{w}_{n}$). Note that the arguments
used for obtaining Eq.~(\ref{tunnel}) are also valid for MTJs with
two ferromagnetic electrodes whose magnetizations are parallel to
each other along the direction $\hat{\mathbf{m}}$. The expansion
coefficients $a_{in}^{(j)}$ ($i=1,2;\;j=0,1,2$) refer to the
system in the absence of the SOC field and therefore do not depend
on $\hat{\mathbf{m}}$. Since these coefficients reflect the cubic
symmetry of the involved bulk materials they obey the relations
$a_{in}^{(j)}(k_{x},k_{y})=a_{in}^{(j)}(-k_{x},-k_{y})$,
$a_{in}^{(j)}(k_{x},k_{y})=a_{in}^{(j)}(-k_{x},k_{y})$, and
$a_{in}^{(j)}(k_{x},k_{y})=a_{in}^{(j)}(k_{y},k_{x})$. Cases in
which the involved materials have other than cubic symmetry in
their bulk form can be treated analogously.

Within linear response theory, the conductance $G$ through the MTJ
is determined by the states at the Fermi energy $E_{F}$ and the
$\mathbf{k}$ dependence reduces to the in-plane
$\mathbf{k}_{\parallel}$ dependence at $E=E_{F}$. One can then
write
\begin{equation}\label{conduct}
    G(\hat{\mathbf{m}})=\frac{g_{_0}}{8\pi^2}\sum_{n}\langle
    T_{n}(\mathbf{k}_{\parallel},\hat{\mathbf{m}})\rangle,
\end{equation}
where $g_{_0}=2e^2/h$ is the conductance quantum and $\langle
...\rangle$ denotes evaluation at $E=E_{_F}$ and integration over
$\mathbf{k}_{\parallel}$.

The time reversal symmetry implies that
$T_{n}(\mathbf{k},\hat{\mathbf{m}})=T_{n}(-\mathbf{k},-\hat{\mathbf{m}})$
and $\mathbf{w}_{n}(\mathbf{k})=-\mathbf{w}_{n}(-\mathbf{k})$. It
follows then from Eq.~(\ref{tunnel}) that the first-order term in
the expansion must be an odd function of $\mathbf{k}$ and will,
therefore, vanish after integration over $\mathbf{k}_{\parallel}$.
As a result the conductance can be rewritten as
\begin{equation}\label{gtot}
    G(\hat{\mathbf{m}})=G^{(0)}+G^{(2)}_{\rm iso}+G^{(2)}_{\rm aniso}(\hat{\mathbf{m}}),
\end{equation}
where $G^{(0)}$ is the conductance in the absence of SOC,
$G^{(2)}_{\rm iso}\propto\langle
    a_{1n}^{(2)}(\mathbf{k}_{\parallel}|\mathbf{w}_{n}(\mathbf{k}_{\parallel})|^2)\rangle$
and
\begin{equation}\label{g-aniso0}
    G^{(2)}_{\rm aniso}(\hat{\mathbf{m}})=\frac{g_{_0}}{8\pi^2}\sum_{n}\langle
    a_{2n}^{(2)}(\mathbf{k}_{\parallel})[\hat{\mathbf{m}}\cdot\mathbf{w}_{n}(\mathbf{k}_{\parallel})]^{2}\rangle,
\end{equation}
are the isotropic and anisotropic SOC contributions, respectively.
In terms of the components of $\hat{\mathbf{m}}$ and the SOC
field, Eq.~(\ref{g-aniso0}) reduces to
\begin{equation}\label{g-aniso}
    G^{(2)}_{\rm aniso}(\theta,\phi)=\frac{g_{_0}}{8\pi^2}{\rm Tr}[{\rm A}{\rm M}(\theta,\phi)],
\end{equation}
where ${\rm A}$ and ${\rm M}(\theta,\phi)$ are matrices whose
elements are given by
\begin{equation}\label{a-matrix}
    {\rm A}_{ij}=\sum_{n}\langle
    a_{2n}^{(2)}(\mathbf{k}_{\parallel})w_{ni}w_{nj}\rangle\;\;\;\;(i,j=x,y,z)
\end{equation}
and
\begin{equation}\label{m-matrix}
    {\rm M}_{ij}(\theta,\phi)=m_{i}(\theta,\phi)m_{j}(\theta,\phi)\;\;(i,j=x,y,z),
\end{equation}
respectively.

Equations (\ref{g-aniso})-(\ref{m-matrix}) are quite general and
reveal how the symmetry of the SOC field can lead to the
anisotropy of the conductance. Further simplifications of these
expressions can be realized by taking into account the properties
of the specific form of the SOC field. In what follows we focus on
the case in which the components of the SOC field can be written
as
$w_{ni}=\mathbf{d}_{ni}\cdot \mathbf{k}_{\parallel}$.
As shown bellow, many relevant physical situations correspond to
such a case. The matrix elements in Eq.~(\ref{a-matrix}) then
reduce to
\begin{equation}\label{a-matrixN}
    {\rm A}_{ij}=\sum_{n}c_{n} \left(\mathbf{d}_{ni}\cdot\mathbf{d}_{nj}\right).
\end{equation}
In obtaining Eq.~(\ref{a-matrixN}) we took into account the
four-fold symmetry of the expansion coefficients
$a_{2n}^{(2)}(\mathbf{k}_{\parallel})$ from which follows that the
only non-vanishing averages are of the form
$c_{n}=\langle
    a_{2n}^{(2)}(\mathbf{k}_{\parallel})k_{x}^{2}\rangle=\langle
    a_{2n}^{(2)}(\mathbf{k}_{\parallel})k_{y}^{2}\rangle$.
By using the Eqs.~(\ref{g-aniso}), (\ref{m-matrix}), and
(\ref{a-matrixN}) the anisotropic part of the conductance can be
rewritten as
\begin{equation}\label{g-anisoN}
     G^{(2)}_{\rm aniso}=\frac{g_{_0}}{8\pi^2}
     \sum_{i,j,n}c_{n}m_{i}(\theta,\phi)m_{j}(\theta,\phi)
     \left(\mathbf{d}_{ni}\cdot\mathbf{d}_{nj}\right).
\end{equation}

The dependence of the TAMR ratio on the magnetization direction is
determined by the anisotropic part of the conductance. Thus,
Eq.~(\ref{g-anisoN}) is our starting formula for discussing
important particular cases.

\section{Results}

We first neglect the effects of strain and focus on the
particularly relevant case in which the SOC field results from the
interference of the Bychkov-Rashba and Dresselhaus SOCs. Later on
we shall consider also MTJs with SOC induced by uniaxial strain.

The Bychkov-Rashba SOC originates from the structure inversion
asymmetry (SIA) of the junction and is basically determined by the
strong electric fields at the interfaces of the tunneling barrier.
It is present, for example, in MTJs with the left and right
electrodes made of different materials and therefore with broken
inversion symmetry. The Dresselhaus SOC results from the bulk
inversion asymmetry (BIA) of one or more of the constituent
materials. Typical materials with BIA are the zinc blende
semiconductors. Thus, the Dresselhaus SOC can be relevant for MTJs
with non-centrosymmetric semiconductor barriers.

The specific form of the SOC field depends on the growth direction
of the heterostructure. Below we analyze the most relevant cases,
corresponding to MTJs grown in the $[001]$, $[110]$, and $[111]$
crystallographic directions.

\subsection{(001) MTJs with axes $\hat{\mathbf{x}}\parallel [110]$, $\hat{\mathbf{y}}\parallel [\bar{1}10]$,
$\hat{\mathbf{z}}\parallel [001]$}\label{case-a}

In this case the SOC corresponding to the $n$th band containing
both Bychkov-Rashba and Dresselhaus terms is given
by\cite{Bychkov1984:JPC,Dresselhaus1955:PR,Fabian2007:APS}
\begin{equation}\label{br-d-001}
    H_{\rm SO}=(\alpha_{n}-\gamma_{n})k_{x}\sigma_{y}-
    (\alpha_{n}+\gamma_{n})k_{y}\sigma_{x},
\end{equation}
where $\alpha_{n}$ and $\gamma_{n}$ are the corresponding
Bychkov-Rashba and Dresselhaus parameters, respectively.

One can extract the components of the SOC field by comparing
Eqs.(\ref{soi}) and (\ref{br-d-001}). It follows then from
equation (\ref{g-anisoN}) that the angular dependence of the
anisotropic conductance is given by
\begin{equation}\label{g-001}
    G^{(2)}_{\rm aniso}=\frac{g_{_0}\sin^{2}\theta}{8\pi^2}
     \sum_{n}c_{n}\left[\left(\alpha_{n}^2+\gamma_{n}^2\right)+
     2\alpha_{n}\gamma_{n}\cos(2\phi)\right].
\end{equation}
The expression above together with Eqs.~(\ref{tamr-in-def}) and
(\ref{tamr-out-def}) lead to the relations corresponding to case A
in Table \ref{tab}.\cite{note0} The obtained TAMR coefficients,
which are valid up to second order in the SOC field, reveal a
clear distinction between the in-plane and out-of-plane
configurations in [001] MTJs: while for a finite out-of-plane TAMR
the presence of only one of the SOCs suffices (i.e., it is
sufficient to have $\alpha_n \neq 0$ or $\gamma_n \neq 0$), the
two-fold symmetric in-plane TAMR appears because of the
interference of non-vanishing Bychkov-Rashba and Dresselhaus SOCs
(i.e., both $\alpha_n$ and $\gamma_n$ have to be
finite).\cite{note1} This explains why a finite out-of-plane TAMR
appears in MTJs such as Fe(001)/vacuum/Cu(001) in which only the
Bychkov-Rashba SOC is present\cite{Chantis2007:PRL} and is in
agreement with the recent observation of the in-plane TAMR in
epitaxial (001) Fe/GaAs/Au MTJs,\cite{Moser2007:PRL} where due to
the presence of the non-centrosymmetric zinc blende semiconductor
GaAs as the barrier material not only the Bychkov-Rashba but also
the Dresselhaus SOC become relevant. In both the in-plane and
out-of-plane configurations, angular dependencies of the form
${\rm TAMR}^{\rm in}_{[110]}(\phi)\propto [1-\cos(2\phi)]$ and
${\rm TAMR}^{\rm out}_{[110]}(\theta)\propto [\cos(2\theta)-1]$
(see Table \ref{tab}) have been experimentally
measured.\cite{Saito2005:PRL,Gao2007:PRL,Moser2007:PRL,Park2008:PRL}
%

The results displayed in Table \ref{tab} suggest the possibility
of using different configurations and reference axes as
complementary setups for TAMR measurements. In particular, our
theoretical model predicts that in the regime $\alpha_{n}\approx
\gamma_{n}$ the out-of-plane TAMR with reference axis in the
$[\bar{1}10]$ is suppressed, while it remains finite if the
complementary axis $[110]$ is used as a reference.\cite{note2} The
opposite behavior, i.e., ${\rm TAMR}_{[\bar{1}10]}^{\rm out}\neq
0$ and ${\rm TAMR}_{[110]}^{\rm out}=0$, is expected when
$\alpha_{n}\approx -\gamma_{n}$. Another relevant regime occurs
when $\alpha_{n}\approx 0$ for which the in-plane TAMR is expected
to vanish (see case A in Table \ref{tab}). The existence of such a
regime was previously invoked in
Refs.~\onlinecite{Moser2007:PRL,Fabian2007:APS} for explaining the
suppression of the in-plane TAMR experimentally observed in (001)
Fe/GaAs/Au MTJs.\cite{Moser2007:PRL} Our theory predicts that
although the in-plane TAMR vanishes, the out-of-plane TAMR should
remain finite in such a regime. In fact, in the regime
$\alpha_{n}\approx 0$ the amplitude of the out-of-plane TAMR
constitutes a direct measurement of the effects of BIA in the
non-centrosymmetric barrier.

\begin{table*}
\caption{TAMR coefficients in units of
$g_{_0}/[16\pi^{2}(G^{(0)}+G^{(2)}_{\rm iso})]$ for different
structures, reference axes $[x]$, and configurations.} \label{tab}
\begin{ruledtabular}
\begin{tabular}{ccccc}
  Case & Structure & $[x]$ & in-plane ${\rm TAMR}^{\rm in}_{[x]}(\phi)$ &
  out-of-plane ${\rm TAMR}^{\rm out}_{[x]}(\theta)$ \\
  \hline
  A & (001) MTJ & $[110]$ & $4[1-\cos(2\phi)]
    \sum_{n}c_{n}\alpha_{n}\gamma_{n}$ &
    $[\cos(2\theta)-1]
    \sum_{n}c_{n}(\alpha_{n}+\gamma_{n})^2$\\
    & & $[\bar{1}10]$ & $4[\cos(2\phi)-1]
    \sum_{n}c_{n}\alpha_{n}\gamma_{n}$ &
    $[\cos(2\theta)-1]
    \sum_{n}c_{n}(\alpha_{n}-\gamma_{n})^2$\\
  \hline
  B & (110) MTJ & $[\bar{1}10]$ & $[1-\cos(2\phi)]\sum_{n}c_{n}(\alpha_{n}^{2}-\beta_{n}^{2})$ &
    $[\cos(2\theta)-1]
    \sum_{n}c_{n}(\alpha^{2}_{n}-\lambda^{2}_{n})$ \\
    & & $[001]$ & $[\cos(2\phi)-1]\sum_{n}c_{n}(\alpha_{n}^{2}-\beta_{n}^{2})$ &
    $2
    \sum_{n}c_{n}[\lambda^{2}_{n}-(\beta_{n}\sin\theta+\lambda_{n}\cos\theta)^2]$ \\
  \hline
  C & (111) MTJ & $[11\bar{2}]$& 0 &
  $[\cos(2\theta)-1]
   \sum_{n}c_{n}(\alpha_{n}+\gamma_{n})^2$ \\
   & & $[\bar{1}10]$ & 0 & $[\cos(2\theta)-1]
   \sum_{n}c_{n}(\alpha_{n}+\gamma_{n})^2$ \\
  \hline
  D & strained (001) MTJ & $[110]$ & $4[1-\cos(2\phi)]
    \sum_{n}c_{n}\eta_{n}\mu_{n}$ &
    $[\cos(2\theta)-1]
    \sum_{n}c_{n}(\eta_{n}+\mu_{n})^2$\\
    & & $[\bar{1}10]$ & $4[\cos(2\phi)-1]
    \sum_{n}c_{n}\eta_{n}\mu_{n}$ &
    $[\cos(2\theta)-1]
    \sum_{n}c_{n}(\eta_{n}-\mu_{n})^2$\\
\end{tabular}
\end{ruledtabular}
\end{table*}

By combining the results shown in Table \ref{tab} one can find
expressions such as
\begin{equation}\label{corre}
    {\rm TAMR}_{[110]}^{\rm in}(90^{\circ})={\rm TAMR}_{[\bar{1}10]}^{\rm
    out}(90^{\circ})-{\rm TAMR}_{[110]}^{\rm
    out}(90^{\circ}),
\end{equation}
which correlate the in-plane and out-of-plane TAMR coefficients
and can be experimentally tested.

\subsection{(110) MTJs with axes $\hat{\mathbf{x}}
\parallel [\bar{1}10]$, $\hat{\mathbf{y}}\parallel [001]$,
$\hat{\mathbf{z}}\parallel [110]$}\label{case-b}

In this case the SOC is described
by\cite{Cartoixa2006:PRB,Diehl2007:NJP}
\begin{equation}\label{br-d-110}
    H_{\rm SO}=\alpha_{n}k_{y}\sigma_{x}-\beta_{n}k_{x}\sigma_{y}-\lambda_{n}k_{x}\sigma_{z}.
\end{equation}
Here $\alpha_{n}$ and $\beta_{n}$ are parameters related to the
SIA-induced SOC, while $\lambda_{n}$ characterizes the strength of
the SOC resulting from the BIA. Note that because of the reduced
symmetry of the (110) structures with respect to the (001) MTJs,
in the present case the usual SIA-induced SOC acquires, in
addition to the usual Bychkov-Rashba SOC, an extra contribution
which leads to $\alpha_{n}\neq \beta_{n}$ in
Eq.~(\ref{br-d-110}).\cite{Cartoixa2006:PRB,Diehl2007:NJP}

Proceeding in the same way as in Sec.~\ref{case-a} we obtain the
following relation for the anisotropic contribution to the
conductance
\begin{widetext}
\begin{equation}\label{g-110}
  G^{(2)}_{\rm iso}=\frac{g_{_0}}{8\pi^2}
  \sum_{n}c_{n}\left[\left(\alpha_{n}^2\cos^{2}\phi+\beta_{n}^{2}\sin^{2}\phi\right)\sin^{2}\theta+
     \lambda_{n}^2\cos^{2}\theta+\beta_{n}\lambda_{n}\sin\phi\sin(2\theta)\right].
\end{equation}
\end{widetext}
The corresponding TAMR coefficients are given in Table \ref{tab}.
The relations for case B show that the angular dependence of the
TAMR in both the in-plane and out-of-plane are similar to the ones
obtained for the (001) MTJs [compare the cases A and B in Table
\ref{tab}]. However, their physical origin is now different. In
the present case the in-plane TAMR originates from the SIA induced
SOC while the out-of-plane TAMR has contributions arising from
both SIA- and BIA-like SOCs. Thus, our model predicts that in
(110) MTJs it could be possible to observe the TAMR in the two
configurations even if the tunneling barrier is composed of a
centrosymmetric material. Another observation
is that the out-of-plane TAMR with reference axis along the
$[\bar{1}10]$ direction could be suppressed if under some given
conditions the regime $\alpha_n = \pm \lambda_n$ (for the relevant
to transport bands) is realized [see the out-of-plane TAMR in case
B of Table \ref{tab}]. In such a case, however, the out-of-plane
TAMR with $[001]$ as the reference axis should remain finite.

\subsection{(111) MTJs with axes $\hat{\mathbf{x}}
\parallel [11\bar{2}]$, $\hat{\mathbf{y}}\parallel [\bar{1}10]$,
$\hat{\mathbf{z}}\parallel [111]$}\label{case-c}

For this structure the SOC is given by\cite{Cartoixa2006:PRB}
\begin{equation}\label{br-d-111}
    H_{\rm SO}=(\alpha_{n}+\gamma_{n})(k_{x}\sigma_{y}-
    k_{y}\sigma_{x}),
\end{equation}
where $\alpha_n$ and $\gamma_n$ are the parameters characterizing
the strengths of the SIA- and BIA-like SOCs, respectively.

After computing the anisotropic part of the conductance we obtain
\begin{equation}\label{g-111}
    G^{(2)}_{\rm aniso}=\frac{g_{_0}\sin^{2}\theta}{8\pi^2}
     \sum_{n}c_{n}\left(\alpha_{n}+\gamma_{n}\right)^2.
\end{equation}
This relation leads to the TAMR coefficients given in Table
\ref{tab} for the case C).

In the present case the prediction of a vanishing in-plane TAMR is
remarkable. We have checked that even if the cubic in $k$ terms
are included in the SOC field, the in-plane TAMR still vanishes.
This could be used for experimentally exploring the origin of the
TAMR. If a suppression of the in-plane TAMR is experimentally
observed in (111) MTJs, it will be a strong indication that indeed
the mechanism behind the TAMR is the SIA- and/or the BIA-like
SOCs. On the contrary, if no suppression of the in-plane TAMR is
observed, the role of these spin-orbit interactions as the origin
of the TAMR can be questioned.

Another interesting issue is the possibility of reaching the
condition $\alpha_n = -\gamma_n$ (for the bands relevant to
transport), which leads to a vanishing out-of-plane TAMR (if only
the linear in $k$ terms in the SOC field are relevant), in
addition to the above discussed suppression of the TAMR in the
in-plane configuration.

\subsection{Uniaxial strain in $(001)$ MTJs with axes
$\hat{\mathbf{x}}
\parallel [100]$, $\hat{\mathbf{y}}\parallel [010]$,
$\hat{\mathbf{z}}\parallel [001]$}\label{case-d}

In our previous analysis we have disregarded the effects of the
strain induced SOC which could be relevant for structures whose
constituent materials have a sizable mismatch in their lattice
constants. For a (001) MTJ the SOC induced by strain is, in
general, given by\cite{Pikus:1984}
\begin{widetext}
\begin{eqnarray}\label{strain-soc}
     H_{\rm SO}&=&\alpha_{n}\left[(u_{zx}k_{z}-u_{xy}k_{y})\sigma_{x}+(u_{xy}k_{x}-u_{yz}k_{z})\sigma_{y}+
     (u_{yz}k_{y}-u_{zx}k_{x})\sigma_{z}\right] \nonumber \\
     &+&\gamma_{n}\left[k_{x}(u_{yy}-u_{zz})\sigma_{x}+k_{y}(u_{zz}-u_{xx})\sigma_{y}+k_{z}
     (u_{xx}-u_{yy})\sigma_{z}\right],
\end{eqnarray}
\end{widetext}
where $u_{ij}$ are the components of the strain tensor and
$\alpha_n$ and $\gamma_n$ are materials parameters. The SOC in
Eq.~(\ref{strain-soc}) is quite rich and suggests the possibility
of engineering the strain (see, for example,
Ref.~\onlinecite{DeRanieri2008:NJP}) in order to manipulate the
behavior of the TAMR. Here we do not consider all the
possibilities but focus, for the sake of illustration, on the case
of an in-plane uniaxial strain such that the only non-vanishing
components of the strain tensor are $u_{xx}=u_{yy}\neq
u_{xy}=u_{yx}$. The existence of a similar strain was initially
assumed for explaining TAMR experiments in (Ga,Mn)As/AlOx/Au
MTJs.\cite{Gould2004:PRL,Ruster2005:PRL} For the in-plane uniaxial
strain Eq.~(\ref{strain-soc}) reduces to
\begin{equation}\label{strain-soc-u}
    H_{\rm SO}=\eta_{n}(k_x \sigma_y -k_y \sigma_x)+\mu_{n}(k_x
    \sigma_x-k_{y}\sigma_y),
\end{equation}
where we have introduced the strain-renormalized SIA  and BIA
parameters $\eta_n =\alpha_n u_{xy}$ and $\mu_n =\gamma_n u_{xx}
$, respectively. The corresponding anisotropic contribution to the
conductance is then given by
\begin{equation}\label{g-strain}
    G^{(2)}_{\rm aniso}=\frac{g_{_0}\sin^{2}\theta}{8\pi^2}
     \sum_{n}c_{n}\left[\left(\eta_{n}^2+\mu_{n}^2\right)+
     2\eta_{n}\mu_{n}\sin(2\phi)\right].
\end{equation}

We note that Eq.~(\ref{strain-soc-u}) has the form of interfering
Bychkov-Rashba  and Dresselhaus SOCs in (001) structures with
$\hat{\mathbf{x}}\parallel
[100]$.\cite{Fabian2007:APS,Cartoixa2006:PRB} Therefore,
Eqs.~(\ref{g-001}) and (\ref{g-strain}) are similar. The angle
$\phi$ in Eq.~(\ref{g-001}) is measured with respect to the
crystallographic direction [110] while in Eq.~(\ref{g-strain}) it
is defined with respect to the [100] axis. Thus, by making the
transformation $\phi \rightarrow \phi+\pi/4$ in
Eq.~(\ref{g-strain}) one recovers a relation similar to
Eq.~(\ref{g-001}). Consequently, assuming the direction [110] as
the reference axis for measuring the magnetization direction the
results for the in-plane and out-of-plane TAMR coefficients in
(001) MTJs with in-plane uniaxial strain [see case D in Table
\ref{tab}] are essentially the same as in the case discussed in
\ref{case-a} but with renormalized spin-orbit parameters which now
account for the strain effects.

In our investigation we have assumed specific well-known forms for
the SOC field. For some systems, however, the form of the SOC
field may not be \textit{a priori} known. In such a case one could
use Eqs.~(\ref{g-aniso})-(\ref{m-matrix}) (which are general) and
contrast them with complementary TAMR measurements in both the
in-plane and out-of-plane configurations in order to deduce the
symmetry properties of the SOC fields.

All the calculated TAMR coefficients, if not zero, show a two-fold
symmetry in the $(\theta,\phi)$-space, which is the symmetry that
has been observed in the
experiments.\cite{Saito2005:PRL,Gao2007:PRL,Ciorga2007:NJP,Moser2007:PRL,Park2008:PRL}
Our results are valid up to the second order in the SOC field. In
particular, our predictions for vanishing TAMR under certain
conditions, may change when higher orders in the SOC field become
relevant. The next higher order contributions in the expansion in
Eq.~(\ref{tunnel}) which do not vanish after averaging are those
containing the fourth order in the SOC field terms and terms of
the fourth order in the cosine directions of $\hat{\mathbf{m}}$
which describe the four-fold symmetry inherent to the involved
bulk ferromagnet. These fourth order terms lead to four-fold
symmetric corrections to the TAMR which may be finite even for
(001) MTJs with centrosymmetric barriers for which the second
order in-plane TAMR calculated here vanishes. Although the
two-fold character of the TAMR is, in general unchanged by these
corrections, they may influence the shape of its angular
dependence. Thus, for the kind of systems here considered (see
Table \ref{tab}) any deviation from the $8-like$ polar shape of
the TAMR [see, for example, Fig.2 in
Ref.~\onlinecite{Moser2007:PRL}] is interpreted in our theory as a
manifestation of higher order contributions and/or strain
effects.\cite{note3} Deviations from the $8-like$ polar shape of
the TAMR have been experimentally
observed.\cite{Gao2007:PRL,Ciorga2007:NJP} In fact, it has been
shown that these deviations may appear by increasing the bias
voltage,\cite{Gao2007:PRL} which within the present approach can
be seen as an indication of higher order in the SOC field terms
turning relevant at sufficiently high bias.

In all the above discussions small magnetic fields with negligible
orbital effects were assumed. It has recently been observed in
Fe/GaAs/Au MTJs that for high magnetic fields the orbital effects
do influence the in-plane TAMR.\cite{Wimmer2009} A version of the
phenological model here presented which incorporates the orbital
effects has recently been developed to qualitatively explain the
magnetic field dependence of the in-plane TAMR experimentally
observed in (001) Fe/GaAs/Au MTJs.\cite{Wimmer2009}

\section{Summary}

We formulated a theoretical model in which the way the TAMR
depends on the magnetization orientation of the ferromagnetic
electrode in MTJs is determined by the specific form and symmetry
properties of the interface-induced SOC field. By using the
proposed model, we deduced the angular dependence of the TAMR for
various systems in dependence of their symmetry under spatial
inversion and their growth direction. The effects of in-plane
uniaxial strain were also investigated.

\acknowledgments{We thank S. D. Ganichev, S.~A.~Tarasenko, and D.
Weiss for useful discussions. This work was supported by the
Deutsche Forschungsgemeinschaft via SFB 689.}



\begin{thebibliography}{26}
\bibitem[{not({\natexlab{a}})}]{note1}
\bibinfo{note1}{Here a finite TAMR is still possible if higher order terms in the SOC field are relevant.}
\bibitem[{not({\natexlab{a}})}]{note2}
\bibinfo{note2}{Note that for the out-of-plane configuration
changing the reference axis implies changing the plane in which
the magnetization is rotated [see Fig.~\ref{syst}].}
\bibitem[{not({\natexlab{a}})}]{note3}
\bibinfo{note1}{For the in-plane uniaxial strain considered here [see Eq.~(\ref{strain-soc-u})]
the angular dependence of the in-plane TAMR deviates from the
8-like polar shape only if higher than second order terms in the
SOC field become relevant. However, for other kinds of strain
different shapes may appear already at the second orther in the
SOC field.}
\end{thebibliography}


%



\end{document}